\documentclass[12pt,preprint]{aastex}

\usepackage{natbib}
\newcommand \micronm {\,\mu{\rm m} }
\newcommand \degree {^{\circ} }

\shorttitle{A Candidate for the Most Luminous OB Association in the
  Galaxy} \shortauthors{Rahman, Matzner \& Moon}

\begin{document}
\title{A Candidate for the Most Luminous OB Association in the Galaxy}
\author{Mubdi Rahman, Christopher Matzner \& Dae-Sik Moon}
\affil{Department of Astronomy \& Astrophysics, University of Toronto,
  50 St.~George Street, Toronto, Ontario, M5S 3H4, Canada;}
\email{rahman@astro.utoronto.ca}

\begin{abstract}

  The Milky Way harbors giant \ion{H}{2} regions which may be powered
  by star complexes more luminous than any Galactic OB association
  known. Being across the disk of the Galaxy, however, these brightest
  associations are severely extinguished and confused.  We present a
  search for one such association toward the most luminous \ion{H}{2}
  region in the recent catalog by Murray and Rahman, which, at
  $\sim$9.7 kpc, has recombination rate of $\sim7 \times 10^{51}$
  sec$^{-1}$.  Prior searches have identified only small scale
  clustering around the rim of this shell-like region, but the primary
  association has not previously been identified.  We apply a
  near-infrared color selection and find an overdensity of point
  sources toward its southern central part. The colors and magnitudes
  of these excess sources are consistent with O- and early B-type
  stars at extinctions $0.96 < A_{K} < 1.2$, and they are sufficiently
  numerous ($406\pm102$ after subtraction of field sources) to ionize
  the surrounding \ion{H}{2} region, making this a candidate for the
  most luminous OB association in the Galaxy. We reject an alternate
  theory, in which the apparent excess is caused by localized
  extinction, as inconsistent with source demographics.
\end{abstract}


\section{Introduction} \label{S:Intro}

Very massive OB associations are objects of intense interest.  Given
that the association birthrate ($\dot{\cal N}_{\rm cl}$) is observed
to vary with association mass ($M_{\rm cl}$) as $d\dot{\cal N}_{\rm
  cl}/d\ln M_{\rm cl}\propto1/M_{\rm cl}$ \citep{mckee97}, or possibly
slightly flatter, a large portion of each galaxy's star formation
occurs within its most massive OB associations. Giant OB associations
are most capable of disrupting their gaseous environment, from their
natal molecular clouds to the entire galactic neighborhood, and they
inflate superbubbles which erupt from the disk, feeding a galactic
halo or fountain.  They sculpt and illuminate the giant \ion{H}{2}
regions seen in distant galaxies and are a notable feature of
starburst evolution. Being the most extreme examples of intense star
formation in the current universe, they are laboratories for the
physics of star cluster formation such as environmental influences on
the initial mass function, formation of very massive stars, and
dynamical evolution with rapid stellar evolution.

Although several of the brightest known OB associations within the
Milky Way approach `super star cluster' status
\citep[$M>10^4M_{\sun}$,][]{port2010}, it is possible that even more
luminous associations have escaped detection because of severe
extinction and confusion in the Galactic plane.  For instance, the
Galactic OB associations with the largest known ionizing output, $S$,
are those powering the NGC 3603 \ion{H}{2}, Arches, Quintuplet, and
Galactic Center regions ($\log S/$s$^{-1}\simeq51.5,51.0,50.9,$ and
$50.5$, respectively: \citealt{figer08, conti04}); however, the upper
limit of the Milky Way association distribution is estimated to be
even brighter \citep[$\log S_u/$s$^{-1}\simeq51.7$;][]{mckee97}.

\begin{figure}
\begin{center}
\includegraphics[scale=0.8]{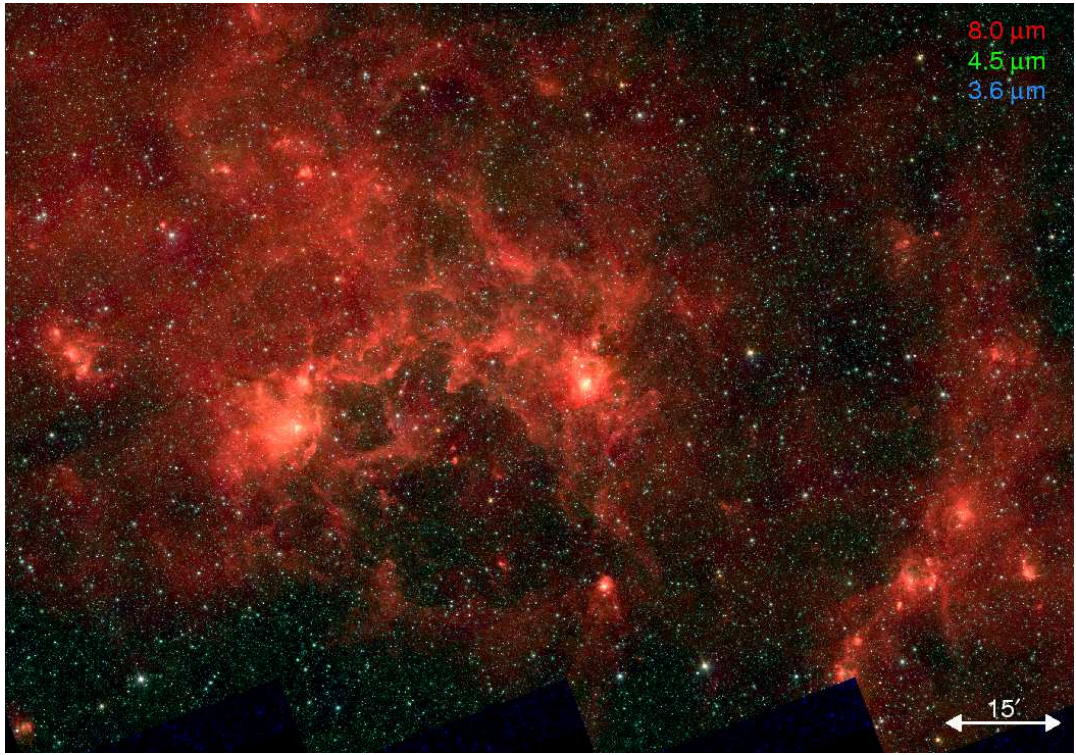}
\end{center}
\caption{G298 or the Dragonfish region seen in the Spitzer GLIMPSE 3.6
  (blue), 4.5 (green) and 8.0 (red) $\micronm$ bands
  \citep{benjamin03}. \label{rgbfig}}
\end{figure}

Recently, \citet{rahman10} identified 40 star-forming complexes within
the 13 most luminous ($S>3\times10^{51}$ s$^{-1}$) free-free emission
regions observed by WMAP and catalogued by \citet{murray10}. Many of
these complexes are potential hosts to embedded super star clusters.
Of particular interest is the most luminous unconfused star-forming
complex located at ($l$, $b$) = ($298.4\degree$, $-0.4\degree$) with a
kinematic distance of 9.7 kpc.  With an output of $\log S/$s$^{-1}
\simeq51.8$, the stars powering this region may represent the Milky
Way's most luminous OB association.  (We refer to this region and its
cluster as G298 or, inspired by its 8\,$\mu$m appearance, as the
`Dragonfish Nebula': Figure \ref{rgbfig}.) The region appears in
8\,$\mu$m as a closed bubble surrounded by a prominent shell.
Previous searches using near-infrared (NIR) point source catalogs from
2MASS \citep{dutra03} and Spitzer GLIMPSE \citep{mercer05} have
concentrated on the brightest regions of free-free emission around the
shell and relied on integrated source counts.  Although some
clustering was discovered in these studies (Figure
\ref{Fig:stellardens}), the central ionizing association is yet
undiscovered.

In this Letter, we identify a candidate for the central OB association
powering the Dragonfish, primarily by applying a color selection to
the 2MASS catalogue.  Our candidate is potentially the most luminous
(and thus massive) OB association in the present Milky Way.

\section{Feasibility and Methods} \label{S:Feasibility} 

Only very luminous stars of deeply embedded OB associations are
visible due to distance and extinction. Whether the association can be
identified using the 2MASS point source catalogue \citep{skrutskie06}
depends on the OB association radius and field star density. 

Radii in the range 10--30\,pc are likely since most associations,
irrespective of mass, are disrupted immediately after birth
\citep{roberts57}.  OB association birthsites in the Milky Way have a
limited range of mass column densities,
$\Sigma\simeq0.3$\,g\,cm$^{-2}\pm0.5\,\mbox{dex}$
\citep{fall10}. Turbulent linewidths in these birthsites are
$\sigma=5\alpha^{1/2}[(0.5/\varepsilon)(\Sigma/.3$\,g\,cm$^{-2})M_5]^{1/4}$
km\,s$^{-1}$, for associations of mass $10^5\,M_5\,M_\odot$ which form
with efficiency $\varepsilon$ from a region with turbulent virial
parameter $\alpha$. Crossing and formation timescales are both shorter
than the ionizing lifetime and the terminal expansion speed of an OB
association is likely to be $\sim\sigma$. For an association of
$\sim10^5M_\odot$, an age between 2--4 Myr, and an expansion speed
between 4--6 km\,s$^{-1}$, we therefore expect an expanding population
currently 8--24\,pc in radius, possibly surrounding a dense core which
survived the disruption event.  For a physical radius $R=10(R/10{\rm
  pc})$ pc with the ionizing luminosity of the Dragonfish as measured
from WMAP, the association source density is $10(R/10{\rm pc})^{-2}$
arcmin$^{-2}$. We measure the 2MASS source population density in the
Galactic plane to be $14-17$ arcmin$^{-2}$. For $R<8$\,pc the
association dominates raw 2MASS source counts, whereas for $R>10$\,pc
it is a minor variation of the 2MASS population which is difficult to
detect using raw source counts.

The NIR color information provides an additional avenue to search for
the embedded associations. We make use of this in the search for the
embedded association within the Dragonfish. Ionization from a coeval
stellar population drops sharply after 4 Myr as its most massive and
powerful stars turn off the main sequence, so it is unlikely, though
not impossible, that G298's ionizing population is any older.  The
ionizing population is sufficiently rich
\citep[$S>10^{50}$\,s$^{-1}$:][]{kennicutt89} to fully sample the
stellar initial mass function (IMF); its very massive stars should
still be present and visible.

The association is likely highly extinguished, even if it has cleared
its natal cloud.  At a distance of 9.7 kpc the fiducial
distance-extinction relationship within the Galactic disk,
$A_V/D\simeq1.6\,\textrm{kpc}^{-1}$ \citep{binney98}, predicts
$A_V\simeq16$; this corresponds to a $K_s$-band extinction
$A_{K_s}\simeq1.0$ \citep{nish08}. We adopt the \citet{nish09}
extinction law, which is steeper than that of \citet{ccm89}.  We rely
primarily on the 2MASS point source catalog for our analysis, but use
three additional data sets to check for consistency: the USNO-B
catalog of visible magnitudes, the Spitzer GLIMPSE catalog at 3.6 and
4.5\,$\mu$m, and deeper J and H band photometry of a small section of
the candidate.

For our search, we target main-sequence, massive stars, at a distance
of $\sim9.7$ kpc and an extinction of around $A_K\simeq1$. Further,
association members visible in all three bands have spectral types of
roughly B1 or earlier, so their intrinsic emission in the NIR bands
($\lambda \sim 1.0-2.4 \,\mu$m) is entirely in the Rayleigh-Jeans
limit \citep[$(J-H,H-K)=(-0.11,-0.10)$;][]{martins06}.  With the
predicted extinction, NIR colors should be $(J-H,H-K)\simeq(1.1,0.5)$.
Association sources should lie in a tight group along the reddening
vector from this common initial color. If the upper IMF follows the
\citet{salpeter55} slope to an upper limit of 120 $M_\odot$, we expect
$\sim400\times10^{-0.34(A_K-1.0)}(\textrm{D}/9.7\textrm{ kpc})^{1.16}$
members with confident NIR colors ($[J,H,K_{\rm s}]<[15.9,15.0,14.3]$,
for which the signal-to-noise ratio exceeds five in all bands).  Note
that if the association is older than 4\,Myr or is not coeval, it must
be even more numerous because more stars are required to make up for
the missing ionization from the top of the main sequence.

The first steps of our analysis are to select sources with the color
of an early-type star behind a range of extinctions bracketing the
expected value, to test for a statistically significant overdensity
within the confines of the 8\,$\mu$m bubble, and then to vary the
adopted extinction range in order to optimize the statistical
significance of the overdensity.  In practice we determine source
density using the angular offset $\theta_N$ between a source and its
$N$th neighbour, as $\Sigma_*=N/(\pi\theta_N^2)$, and use
$N\sim20-100$ to reduce Poisson noise at some expense in
resolution. Once the angular scale of the putative association is
determined, we assess its significance by comparing source density
within this region to the mean $\bar\Sigma_*$ and standard deviation
$\sigma_\Sigma$ of those within independent, identically-sized regions
in a 2-degree field centered on, but excluding, the 8\,$\mu$m
bubble. Statistical significance is measured by
$(\Sigma_*-\bar\Sigma_*)/\sigma_\Sigma$.

Once a candidate is identified, we check that the same procedure
applied to regions outside the 8\,$\mu$m bubble yields no candidates
of similar significance. We then examine the color-color and
color-magnitude diagrams (CCDs and CMDs) of similarly-sized regions to
check that an apparent overdensity is not caused by features of the
distribution of stars and extinction.

\section{Results}\label{Results} 
\subsection{Candidate OB Association} \label{Results:Candidate}

\begin{figure}
  \begin{center}
    \includegraphics[scale=1.0]{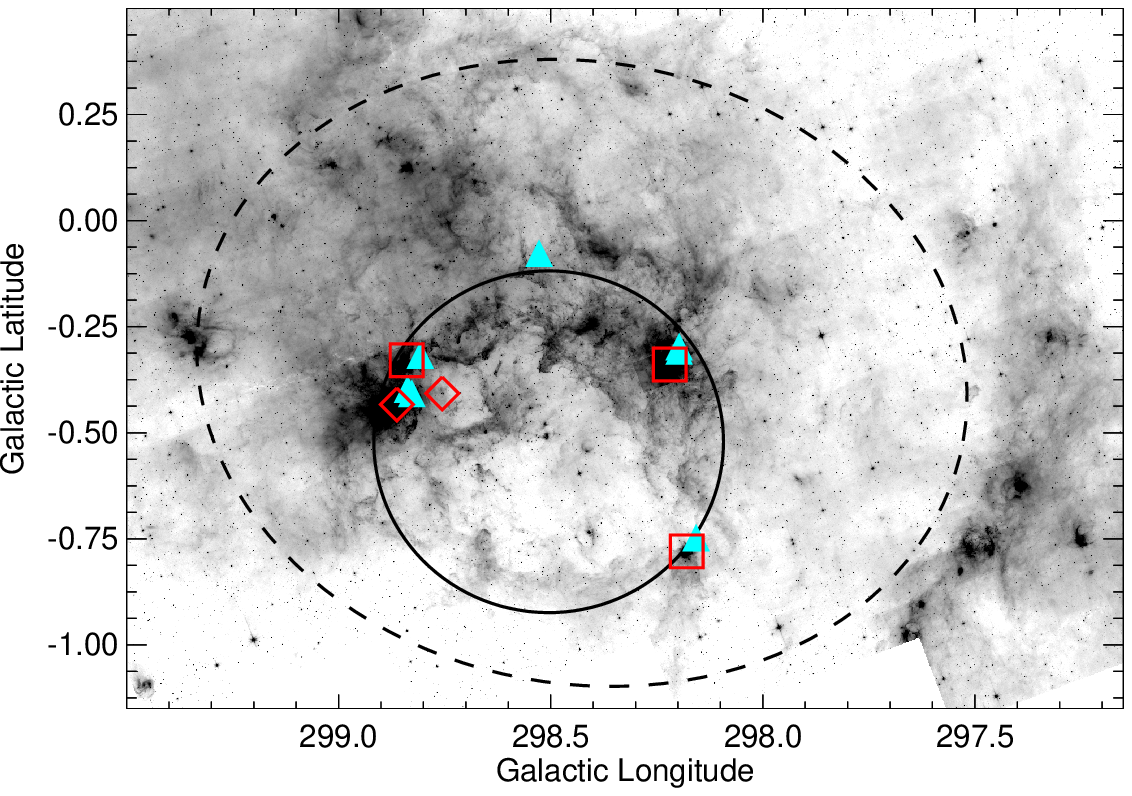}\\
    \includegraphics[scale=0.45]{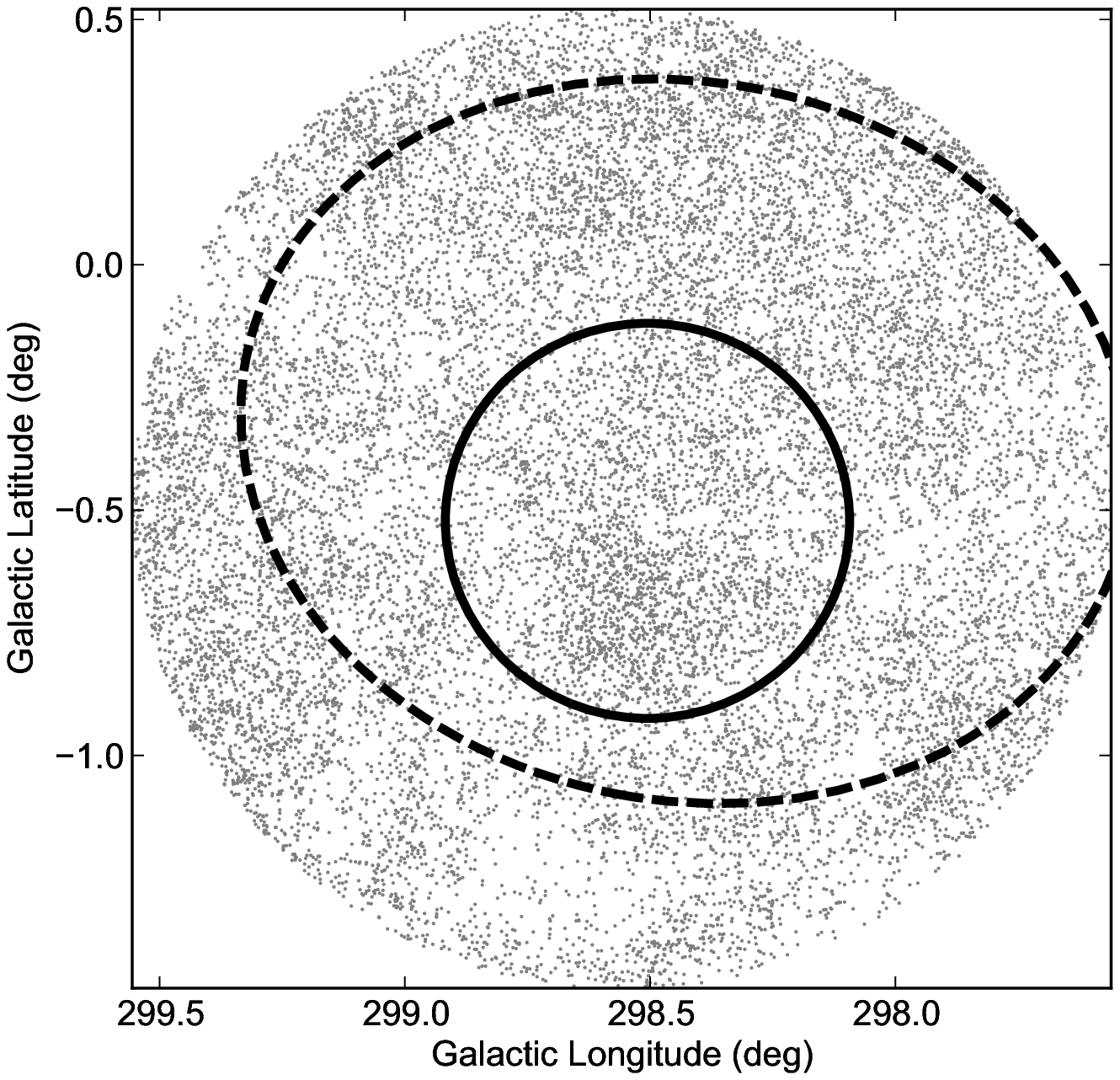}
    \includegraphics[scale=0.45]{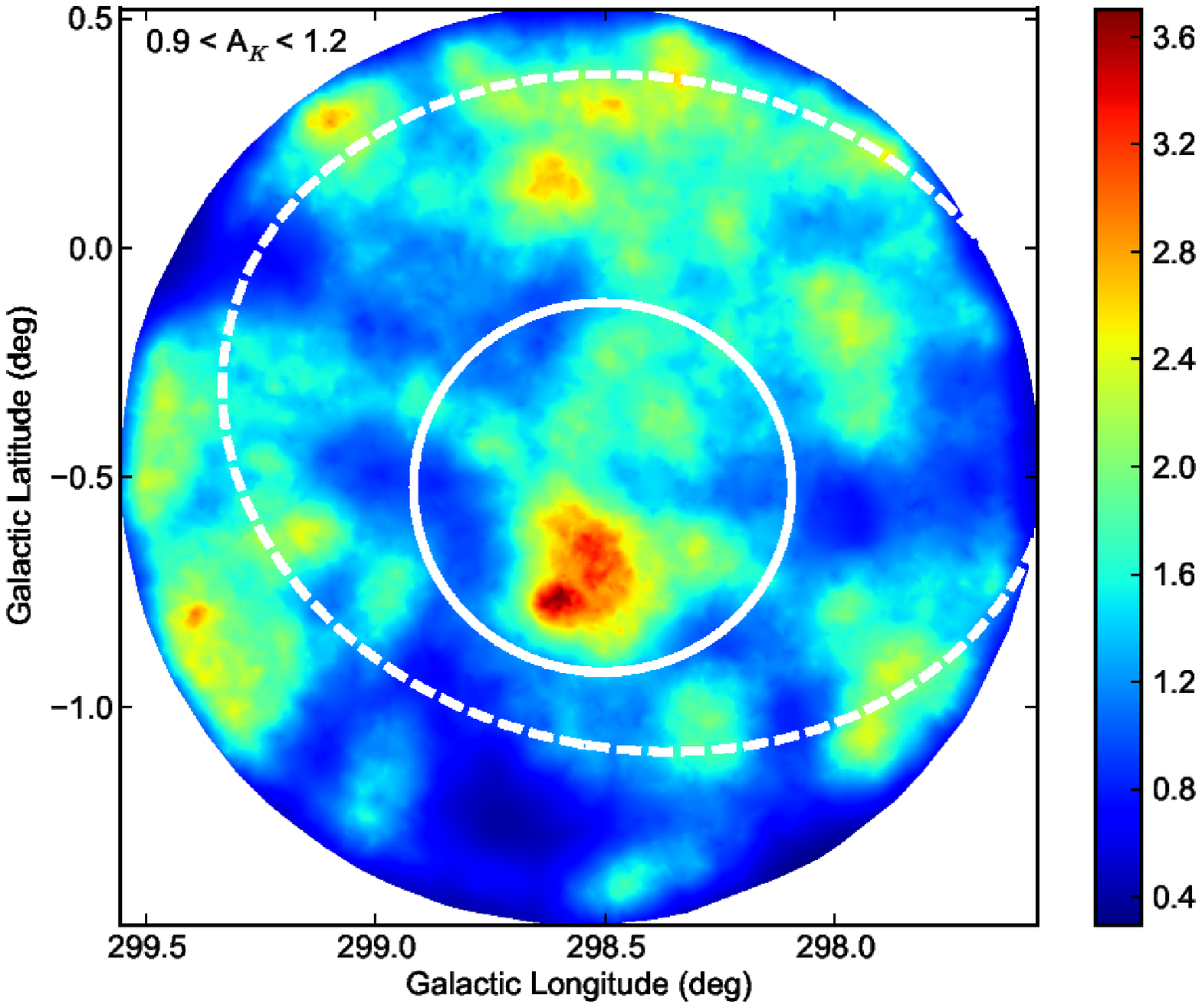} 
  \end{center}
  \caption{Top: The 8$\micronm$ GLIMPSE image of the Dragonfish
    Nebula. The dotted ellipse indicates the location of the WMAP
    source, and the solid ellipse indicates the location of the
    star-forming complex. The cyan triangles indicate the location of
    known \ion{H}{2} regions. Clockwise from top: G298.56-0.11,
    G298.2-0.3, G298.19-0.78, G298.9-0.4, and G298.837-0.347. The red
    squares and diamonds indicate the regions of stellar clustering
    identified by \citet{dutra03} and \citet{mercer05}. Bottom Left:
    The individual 2MASS sources within the field surrounding the
    Dragonfish Nebula passing the color cut. Bottom Right: The 2MASS
    stellar densities of the field surrounding the Dragonfish Nebula
    at the color cut corresponding to O-stars with extinctions between
    $0.9<A_{K}<1.2$. The color bar indicates the stellar density in
    stars per square arcminute.  \label{Fig:stellardens}}
\end{figure}

Based on the expected source colors from \S~\ref{S:Feasibility}, we
apply a NIR color cut of $1.0<J-H<1.4$ and $0.44<H-K_{s}<0.62$,
corresponding to colors of hot stars with $0.9<A_{K}<1.2$. Within the
color cut, the 2-degree diameter field encompassing the Dragonfish has
a mean stellar density $\bar\Sigma_*=1.4$ arcmin$^{-2}$, which we
refer to as the field star density. The standard deviation of
$\Sigma_*$ averaged over association-sized regions is
$\sigma_\Sigma=0.30$ arcmin$^{-2}$. We find a significant overdensity
of stars in the Dragonfish Nebula at ($l$, $b$) = ($298.55\degree$,
$-0.72\degree$), located inside the shell of the star forming complex
(Figure~\ref{Fig:stellardens}).  The 1--$\sigma$ stellar overdensity
contour ($\Sigma_*=\bar\Sigma_*+\sigma_\Sigma$) has semimajor and
semiminor axes (11\arcmin, 10\arcmin), corresponding to (31, 28)pc;
however, the association may extend beyond the observational
limit. The peak of the overdensity is $\Sigma_*=3.7$ arcmin$^{-2}$,
8-$\sigma$ above the background level on cluster-sized regions,
adopting the 100th nearest neighbor. The region contains 897 sources
within the adopted boundaries.  Subtracting the contribution due to
background sources, the candidate association is composed of
$406\pm102$ sources, consistent with the value determined based on the
chosen IMF and measured luminosity (\S~\ref{S:Feasibility}). Using the
1--$\sigma$ boundaries, the mean density of the association is
$\Sigma_*+4.0\sigma_\Sigma$. We refer to the candidate as the
``Dragonfish Association''.

The association is strongly concentrated within the central $4.1'$,
but contains an extended asymmetric envelope. The central, dense,
symmetrical region has a stellar density which depends as
$\theta^{-\alpha}$ with $\alpha = 0.44\pm 0.03$ for $2'<\theta<3'$,
where the value of $\alpha$ depends on the assumed density of field
sources.  At larger radii the azimuthally averaged density drops more
rapidly, but is dominated by the asymmetries visible in Figure 1. Half
the excess sources are found within $4.7'$, or
$13\,(D/9.7\textrm{kpc})$ pc, similar to the half-light radii of other
Galactic OB associations \citep{port2010}.

Assuming half the projected mass falls within this radius, and
assigning a total mass of $10^5\,M_\odot$ we infer that the
association's central regions are older than their virial crossing
times, whereas its outskirts are not ($T_{cr} = 22$ Myr). It therefore
seems likely to be unbound on the whole \citep{gieles11} but to
contain a bound core; however this bears further investigation.

Using the USNO and Spitzer GLIMPSE catalogues, we are able to confirm
that the stars identified in the Dragonfish Association have
magnitudes and colors consistent with OB stars with the stated
extinction, but not further able to distinguish the association
population from the background. In the GLIMPSE wavelengths
($>3\micronm$), this is because nearly all stars have identical,
Rayleigh-Jeans colors, whereas with the USNO wavelengths
($<1\micronm$), the spectral energy distribution is strongly dominated
by the extinction. Thus template colors of most stellar types can be
fit to the additional photometry by varying the assumed distance or
extinction of the star.

\subsection{Color-Color and Color-Magnitude Diagrams }\label{ccsect}

\begin{figure}
  \begin{center}
    \includegraphics[scale=0.70]{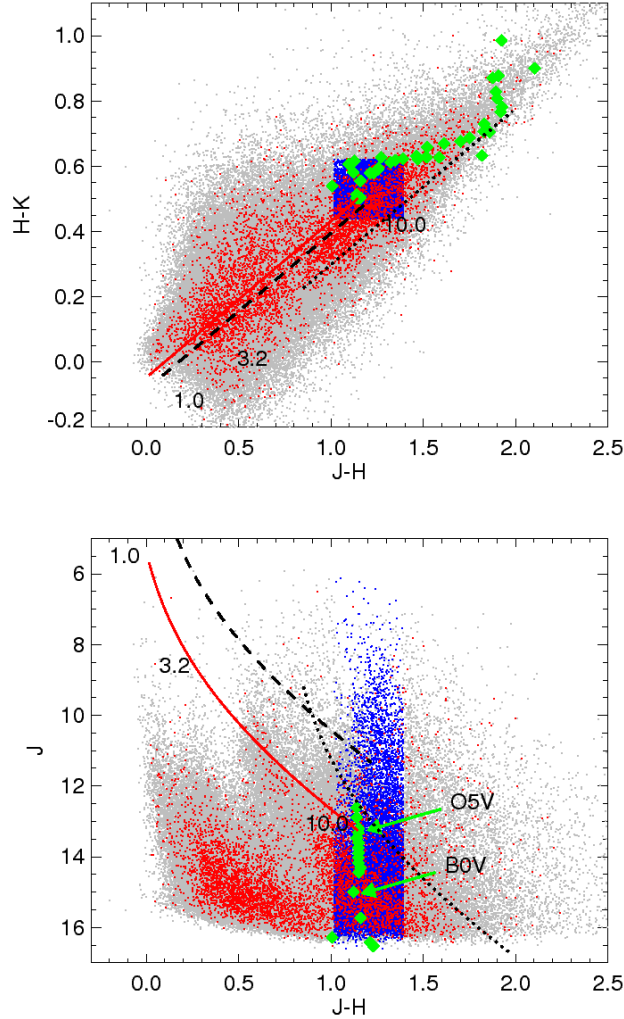}
  \end{center}
  \caption{Color-color (top) and color-magnitude (bottom) diagrams of
    the 2MASS point sources. The grey points are all points within the
    two-degree field, while the red points are the sources within the
    association boundaries. The blue points indicate all sources
    within the color cut. The green diamonds indicate the main
    sequence at a distance of 9.7 kpc with an extinction of
    $A_{K}=1.0$. The location of an O5V and B0V star are indicated for
    reference. The red line indicates the track of an O5V star at
    various distances, with the numbers indicating the distance in
    kpc.  Similar paths are shown for K3II (black dotted line) and B0I
    (black dashed line) stars. \label{cmcc}}
\end{figure}

The NIR color information provides an additional avenue to determine
important parameters of the candidate association, including the
statistical significance of the overdensity, the possible background
contributors, and the color variation of all point sources within the
association boundaries. Figure \ref{cmcc} presents the CCD and CMD of
the point sources within the 2 degree field centered on the Dragonfish
Nebula, showing the distribution of all 2MASS point sources, as well
as the relative location of the color cut. We indicate in red the
point sources that fall within the association boundaries. We indicate
in blue all stars within the field that meet the association's color
cut. The position of the main sequence is indicated at a distance of
9.7 kpc and extinguished by $A_{K}=1.0$. The reference magnitudes for
the O spectral type are taken from \citet{martins06}, and for the
remaining spectral types from \citet{pickles98}. The path taken by an
O5V star if placed at varying distances, assuming the average
extinction to distance ratio, is indicated by the red line. The path
taken by any given reference star is just a translation of this path
on the diagram. The color cut is sufficiently large to permit spectral
classes of A and earlier at the given reddening. However, the
sensitivity of 2MASS limits stars at this distance and extinction to
be no later than B1V.

\begin{figure}
  \begin{center}
  \includegraphics[scale=0.8]{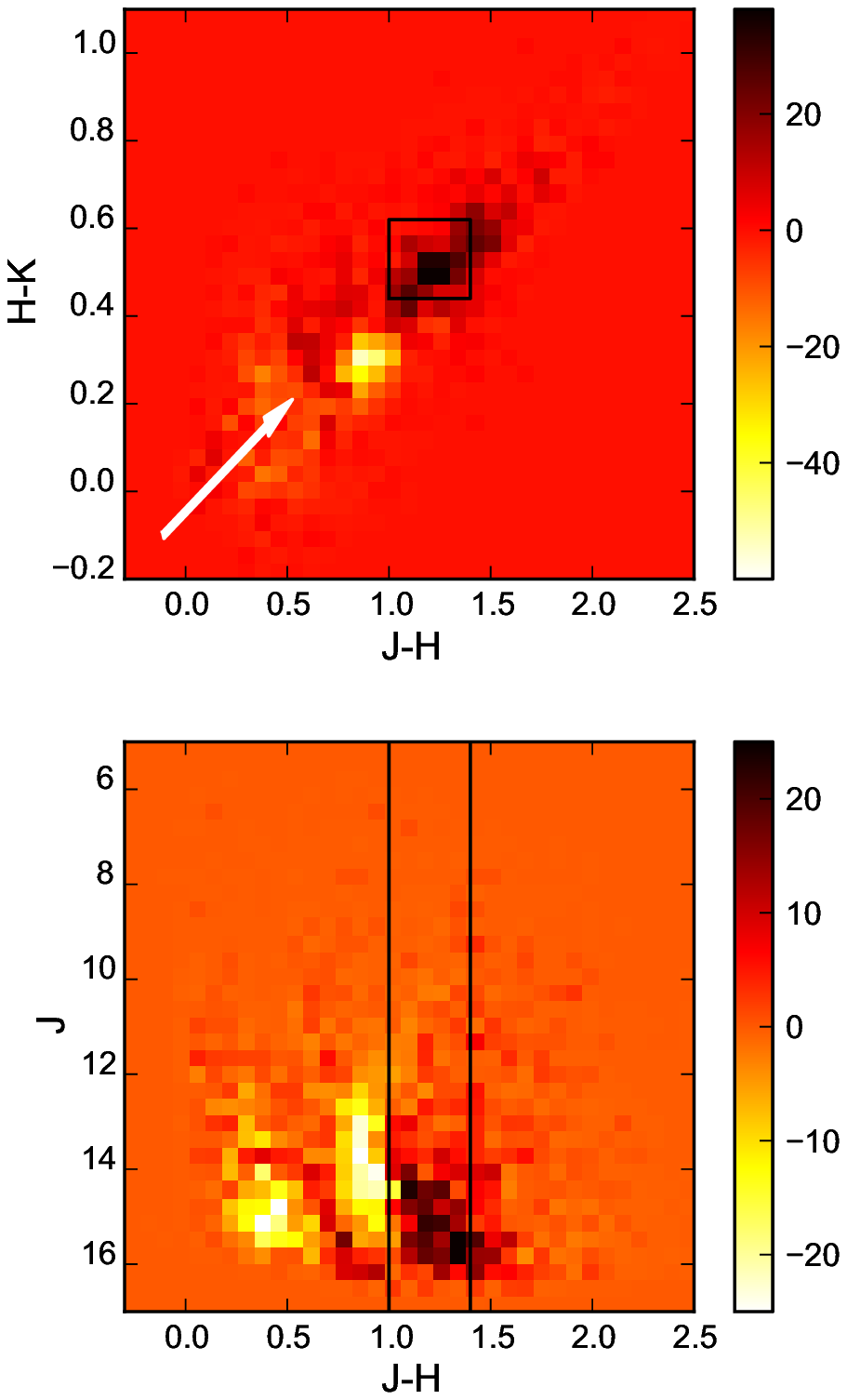}
  \end{center}
  \caption{Binned color-color (top) and color-magnitude (bottom)
    diagrams of the 2MASS point sources within the candidate
    association boundaries with the binned background removed from the
    diagrams. The cluster color cut is indicated with the black
    lines. The color bars indicate the number of stars in each
    bin. The white arrow on the color-color diagram indicates the
    extinction vector corresponding to an A$_{K} = 0.5$ from the
    unreddened location of O stars. \label{binnedcm}}
\end{figure}

In Figure~\ref{binnedcm} we bin the CCD and CMD to investigate the
overabundance in the region within the Dragonfish Association as
compared to the surrounding regions. These figures indicate the
location and prominance of the cluster overdensity as a function of
both color and magnitude. For the CCD, we bin the sources in a $35^2$
bin grid spanning $-0.3<J-H<2.5$ and $-0.2<H-K<1.1$. For the CMD, we
similarly use a $35^2$ bin grid with $5<J-H<17.0$, and identical $J-H$
limits. We grid the point sources that fall within the association
boundaries and subtract the average bin value from the surrounding
field.  In both the CCD and CMD, an overabundance of point sources
appears in the range $1.0<J-H<1.5$. We note an underdensity in the CCD
at $(J-H,\,H-K)=(0.8,0.3)$. The total number of sources ``missing'' in
the underdensity is approximately $~200$. The underdensity is
associated with a galactic structure feature in the surrounding field
which we discuss in depth in a forthcoming paper.

Given that the colors of these underdense sources are separated from
those of the putative OB association by a vector which is nearly
parallel to the extinction vector, could the association be only an
illusion caused by intervening extinction with $A_K\sim 0.25$? If so
we would expect an overall deficit of stars in this direction (due to
the extinction of dim sources), but there is no such trend.  In fact
the missing sources are about half as numerous as the surplus ones.
Corroborating this point, we note that the vector separating the two
populations is slightly steeper than the \citet{nish08} extinction law
(whereas the overpopulation is perfectly consistent with OB stars
reddened by that law).  Moreover, sources which appear underdense on
this plot vary on larger angular scales than that of the association;
their population is mostly a function of galactic latitude, although
there are somewhat fewer of them toward the association. We therefore
conclude that the association does in fact exist, but there may also
exist localized extinction which affects the color distribution of
sources.

To further investigate the role of patchy extinction, we have obtained
additional $J$ and $H$ band photometry towards a central 3.9\arcmin{}
field in the candidate association using the Wide Field IR Camera on
the 2.5-metre du Pont Telescope, with limiting magnitudes of $J=17.1$,
and $H=16.3$ at a SNR of 3 (J. Radigan, private communication, 2010).
We extract 906 stars in this field. With this additional photometry,
we find stars in the range of $0.8<J-H<0.96$, with a continuous
density of stars throughout this color range. We also find no evidence
for any shadowing features which would indicate a strong extinction
feature. This supports our conclusion that the apparent association is
not an extinction feature.

\section{Confusing sources } \label{background}

In order to identify possible contaminating stars within the color
cut, we model the NIR colors and magnitudes of template stars of
different spectral types and luminosity classes.  We vary the
distances of the model stars to examine which stars meet the cut above
the 2MASS limiting magnitudes.  We use reference magnitudes,
extinction relationship and the average extinction-to-distance ratio
as in \S\ref{ccsect}.  All main sequence OB stars down to B1V can fall
within the color and magnitude limits. With the 2MASS limiting
magnitudes, assuming the extinction-to-distance ratio, all subtypes of
O stars are visible to 13 kpc.  All main sequence stellar types later
than B1V fall below the magnitude limits. OB giants (down to B9III)
and K1III to M0III giants can fall within the color and magnitude
limits. All supergiant stars can fall within the association color cut
and magnitude limits. We show examples of stars that may fall within
the color and magnitude limits in Figure \ref{cmcc}.  The red-clump
giant feature, composed of early K giant stars \citep{lopez02,
  indebetouw05} partially overlaps the association color cut towards
the faint end, but is not the primary contributor to the background
population.

Using the \citet{robin04} model of the stellar population we produce a
synthetic catalogue of stars. We simulate the field in the direction
and with the size of the Dragonfish Association, using the 2MASS
limiting magnitudes from \S\ref{S:Feasibility}, and a diffuse
extinction law of 1.6 $A_{V}/\textrm{kpc}$. The model background
population underestimates the total number of sources. However, we
find that the majority of stars within the color cut are early K giant
stars with masses between 1 and 3 M$_{\sun}$ and distances between 3.0
and 4.5 kpc; we assume these represent our contaminating sources. 

\section{Conclusions } \label{conc}

We have used the 2MASS point source catalogue to search for the
central OB association within the Dragonfish Nebula by analyzing the
on-sky stellar density with NIR colors consistent with extinguished O
stars. We find an overdensity of point sources within the bubble at an
extinction range of $0.96<A_{K}<1.2$. Comparing the CCD and CMD of the
candidate association with the surrounding field, we confirm the
presence of the overdensity at the chosen color range. We reject an
extinction explanation for the overdensity and infer that the
contaminating star population consists predominantly of K giants 3-4.5
kpc away. Our candidate association contains $406\pm102$ members
visible in 2MASS with magnitudes consistent with O and early-B stars,
similar to the ionizing population expected from the region's
free-free luminosity.  We infer from its assumed mass, radius, and
maximum age that it is likely to be unbound on the whole but may
contain a bound core; however further investigation is necessary to
determine its dynamical state. This is a candidate for the most
luminous (and consequently the most massive) OB Association in the
Galaxy.  If confirmed by upcoming spectroscopic observations, the
Dragonfish Association is an outstanding candidate for multiwavelength
examination, as it is a promising laboratory massive star formation,
stellar dynamics, and feedback processes within the Galactic
environment.

\acknowledgements

We thank J. Radigan for obtaining NIR data for us. We thank N. Murray,
P.G. Martin, and R. Breton for the many helpful discussions. This
publication makes use of data products from the Two Micron All Sky
Survey, which is a joint project of the University of Massachusetts
and the Infrared Processing and Analysis Center/California Institute
of Technology, funded by the National Aeronautics and Space
Administration and the National Science Foundation.

\end{document}